\documentclass[11pt, a4paper, twocolumn]{article}
\usepackage[a4paper, total={6.5in, 9.5in}]{geometry}
\usepackage{xr}
\usepackage{hyperref}
\usepackage{amssymb}
\usepackage{xurl}
\usepackage[backend=biber, style=nature]{biblatex}
\usepackage{abstract}
\usepackage{authblk}
\usepackage{color}

\usepackage{lipsum}
\usepackage{graphicx}
\hypersetup{colorlinks=false, urlcolor=blue, linkcolor=blue, citecolor=blue}

\makeatletter
\newcommand*{\addFileDependency}[1]{
\typeout{(#1)}
\@addtofilelist{#1}
\IfFileExists{#1}{}{\typeout{No file #1.}}
}\makeatother

\newcommand*{\myexternaldocument}[1]{%
\externaldocument{#1}%
\addFileDependency{#1.tex}%
\addFileDependency{#1.aux}%
}
\myexternaldocument{Supplementary_material}

\title{Nano-assembled open quantum dot nanotube devices}

\author[1,a]{Tim Althuon}
\author[2,a,*]{Tino Cubaynes}
\author[1]{Aljoscha Auer}
\author[1]{Christoph Sürgers}
\author[1,2]{Wolfgang Wernsdorfer}
\affil[1]{Physikalisches Institut, Karlsruhe Institute of Technology, 76131 Karlsruhe, Germany}
\affil[2]{Institut für QuantenMaterialien und Technologien, Karlsruhe Institute of Technology, 76344 Eggenstein-Leopoldshafen, Germany}
\affil[a]{These authors contributed equally}
\affil[*]{Corresponding author: tino.cubaynes@kit.edu}

\begin{document}



\twocolumn[
\begin{@twocolumnfalse}
\maketitle
\begin{abstract}
A pristine suspended carbon nanotube is a near ideal environment to host long-lived quantum states. 
For this reason, they have been used as the core element of qubits and to explore numerous condensed matter physics phenomena. One of the most advanced technique to realize complex carbon nanotube based quantum circuits relies on a mechanical integration of the nanotube into the circuit. Despite the high-quality and complexity of the fabricated circuits, the range of possible experiments was limited to the closed quantum dot regime. Here, by engineering a transparent metal-nanotube interface, we developed a technique that overcomes this limitation. We reliably reach the open quantum dot regime as demonstrated by measurements of Fabry-Perot interferences and Kondo physics in multiple devices. A circuit-nanotube alignment precision of $\pm 200\,$nm is demonstrated. Our technique allows to envision experiments requiring the combination of complex circuits and strongly coupled carbon nanotubes such as the realization of carbon nanotube superconducting qubits or flux-mediated optomechanics experiments.
\end{abstract}
\end{@twocolumnfalse}]


\subsection*{Introduction}

Nanocircuits based on carbon nanotubes (CNT) have shown to be a very versatile platform to explore a broad range of quantum phenomena. As a nearly perfect 1D system and a pristine material, suspended carbon nanotubes allowed to realize Wigner 1D crystals \cite{deshpande2008one, pecker2013observation, shapir2019imaging}, spin-qubits \cite{laird2013valley, cubaynes2019highly} or nanomechanics experiments \cite{lassagne2009coupling, steele2009strong, ares2016sensitive, urgell2020cooling, wen2020coherent}, among others. Because the nano-fabrication technique used to build such circuits ultimately limits their complexity, the development of new techniques is a crucial point to envision more advanced experiments. The most delicate step is the integration of the CNT into the circuits while maintaining its cleanliness. The most recently developed approach, referred to as "stamping" technique relies on a mechanical transfer of the CNT into the circuit at the last fabrication step \cite{waissman2013realization, wu2010one, muoth2012transfer, pei2012valley, blien2018quartz, gramich2015fork, cubaynes2020nanoassembly}. In addition to producing ultraclean CNTs \cite{waissman2013realization}, this technique has the strong advantage to be a deterministic process in which a single CNT can be pre-characterized via electrical transport \cite{waissman2013realization} or imaging (this work) and only then integrated at a precise location within a potentially complex circuit. For instance, it allows to produce CNT-based circuits with a record high number of gates (up to 16 gates) \cite{waissman2014carbon}, containing multiple CNTs \cite{waissman2014carbon}. In parallel, the stamping technique establishes compatibility between the CNT technology and more complex systems, making it a particularly powerful technique. This was demonstrated with the integration of a CNT inside a microwave resonator \cite{ranjan2015clean, cubaynes2020nanoassembly, blien2020quantum}.

\begin{figure*}[h!]
\centering
\includegraphics[width=0.85\linewidth]{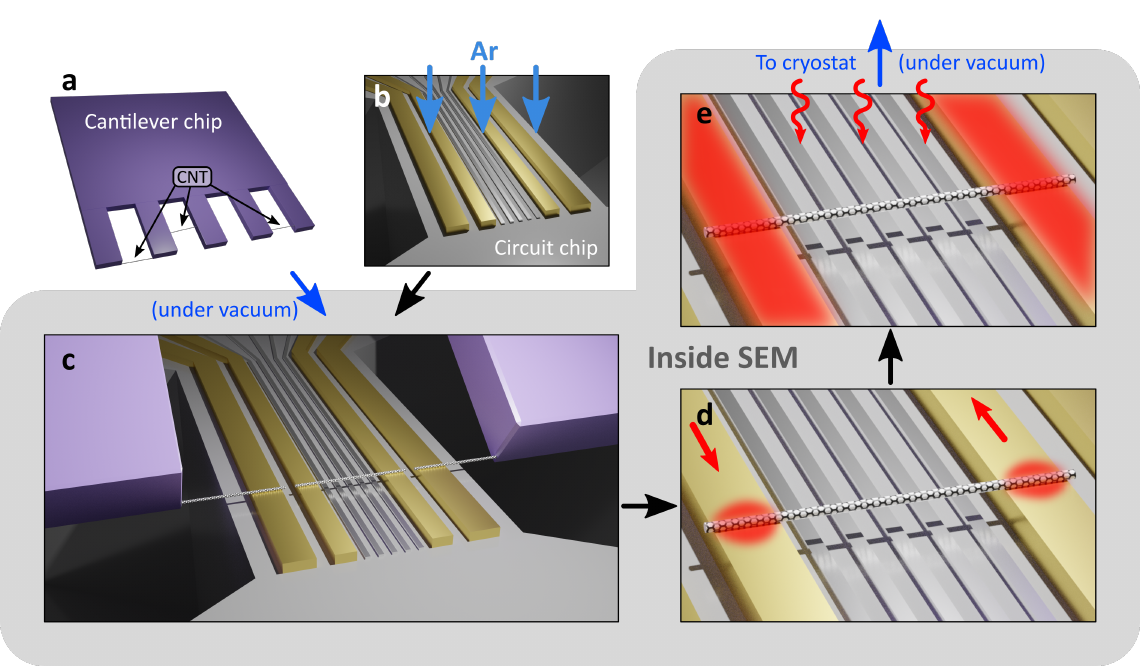}
\caption{\textbf{Nano-assembly and contact improvement strategy.}
\textbf{a.} Simplified representation of the silicon chip on which suspended carbon nanotubes are grown between cantilevers.
\textbf{b.} Typical circuit (here a double quantum dot) onto which a CNT is transferred. An argon-milling is performed on the circuit prior to the loading into the SEM.
\textbf{c.} Nano-assembly step in which the CNT is transferred on the circuit, and disconnected from the cantilevers by driving a high current in the two outermost sections (see methods).
\textbf{d.} Current-induced annealing step. Most of the resistance of the circuit is localized at the CNT-metal interface focussing the heating power directly on the interface as represented with a red halo.
\textbf{e.} Radiative thermal annealing of the circuit using a halogen lamp.
The gray global background indicates the steps performed inside our dedicated SEM.}
\label{fig1}
\end{figure*}

Nevertheless, the stamping approach was until now unable to produce a transparent CNT-metal interface, which leads to device resistances of stamped circuits restricted to $\mathrm{R} \geq 200\,\mathrm{k}\Omega$. As a result, quantum transport measurements have been limited to the closed quantum dot regime epitomized by Coulomb blockade, which constitutes the main restriction of the stamping technique. In the carbon nanotube transistors research community, different methods have been developed to improve the quality of the CNT-metal contact relying on thermal annealing either via high current \cite{woo2007reduced, maki2004local, yang2020fabry} or temperature \cite{jung2021understanding, cao2015end}, but none has been shown to be compatible with the stamping approach.

Besides the stamping technique, another approach which has been very efficient in producing high quality CNT circuits with a transparent CNT-metal interface relies on the growth of the CNT directly on top of the circuit \cite{cao2005electron, steele2009strong, eichler2011nonlinear, tormo2022novel}. Nevertheless, the extreme CVD conditions required to grow CNTs \cite{kong1998synthesis} and the probabilistic nature of this fabrication technique limit it to the fabrication of simple structures (with the exception of a recent work \cite{tormo2022novel}) preventing to envision more advanced experiments.

In this work, we present an approach combining a custom stamping technique with a thorough contact improvement procedure allowing to overcome this limitation. It relies on three key elements: (1) pre-cleaning of the circuit via argon-milling before the transfer of the CNT, (2) maintaining the nanotube under vacuum throughout the full process, and (3) annealing the CNT-metal contact at relatively low temperature (below 300$\,$°C) with a two-step annealing approach combining current-induced and radiative thermal annealing (RTA). We could identify the individual effects of each of them by monitoring the device resistance after each annealing step. 

Using this procedure we obtained device resistances down to $25\,\mathrm{k}\Omega$ at room temperature, hence proving the high transparency of the CNT-metal interface. This low value of  resistance allows to reach the open-quantum dot regime, as demonstrated by the measurements of Kondo ridges and Fabry-Pérot interferences. In addition, our technique allows to position the CNT with an accuracy of $\pm\,$200$\,$nm. The newly accessed open quantum dot regime combined with the flexibility and the accuracy of the stamping nano-fabrication technique, paves the way to more advanced experiments with CNT-based hybrid systems, such as novel superconducting qubits or flux-mediated optomechanics experiments \cite{khosla2018displacemon, rodrigues2019coupling, schmidt2020sideband}.

\begin{figure}[h!]
\centering
\includegraphics[width=0.8\linewidth]{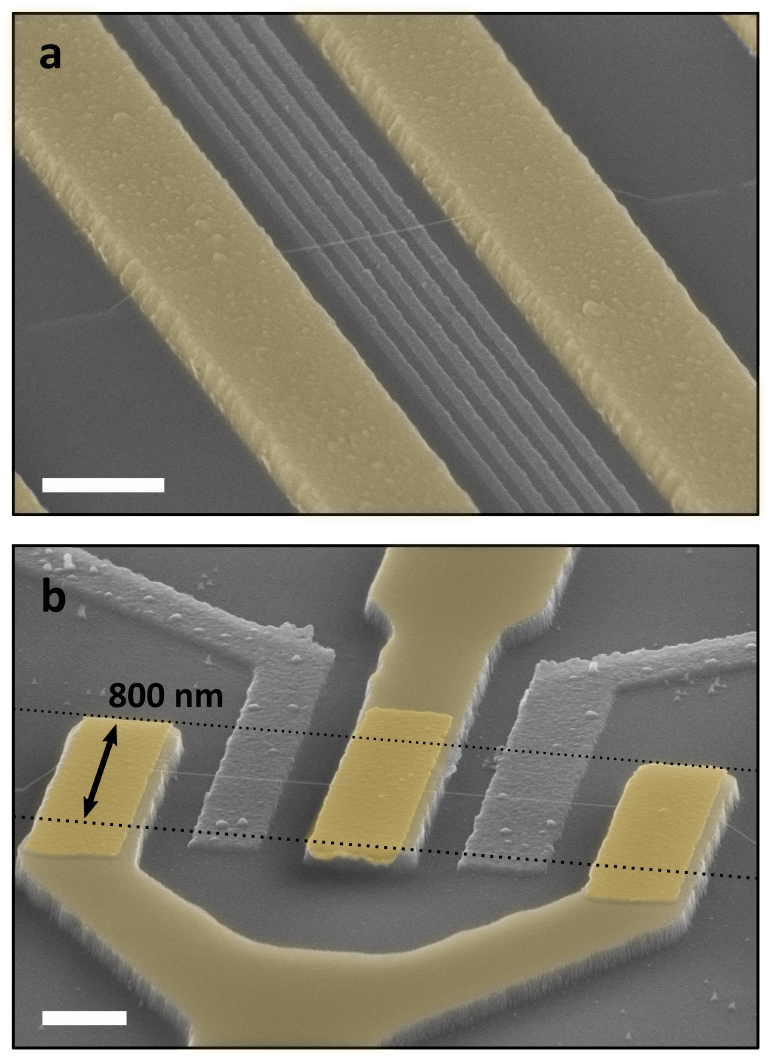}
\caption{\textbf{Examples of nano-assembled devices.}
False color scanning electron micrographs of two different circuits realized with the stamping technique. Electrodes contacting the CNT are highlighted in yellow.
\textbf{a.} A double quantum dot circuit with 5 gates.
\textbf{b.} A circuit with a SQUID geometry (Superconducting Quantum Interference Device). Two dotted lines define a 800$\,$nm-wide region where the CNT must be deposited. The positioning of the CNT in the center of this region allows for a conservative estimation of at most $\pm\,$200$\,$nm deposition accuracy.
Scale bars: 500$\,$nm.}
\label{fig2}
\end{figure}

\begin{figure}[h!]
\centering
\includegraphics[width=1\linewidth]{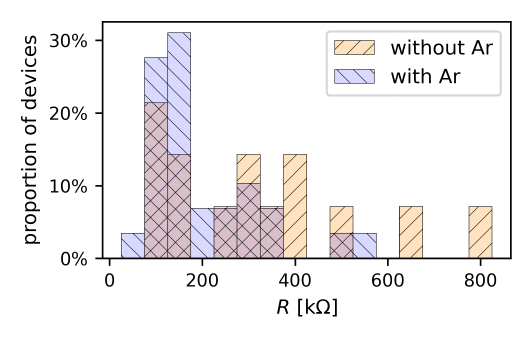}
\caption{\textbf{Influence of argon-milling pre-treatment.}
Device resistance distribution of devices with and without argon-milling (comprising only devices with  R$<$1$\,$M$\Omega$, in total 14 devices without and 29 devices with argon-milling).}
\label{fig3}
\end{figure}

\subsection*{Results and discussion}
\subsubsection*{Pre-cleanings and transfer of the CNT}

Following the principle of the stamping technique, suspended CNTs are grown between individual cantilevers arranged in an array (see Figure \ref{fig1}.a). In parallel, the circuit chip on which a CNT will be transferred is fabricated using standard lithography techniques (see methods). The full device is then assembled by mechanically transferring a single CNT on top of the circuit (details in methods). In Figure \ref{fig2}, two examples of fabricated circuits with different geometries are presented. One specificity of our approach is to perform the deposition of the CNT inside a dedicated Scanning Electron Microscope (SEM), which presents multiple advantages. First, it allows to localize CNTs on the cantilever chip and transfer a single one onto the circuit while maintaining the CNT and circuit chip under vacuum. Secondly, e-beam imaging combined with a custom piezo-motor stage allowed us to position the CNT on the circuit with a precision of $\pm\,200\,$nm (see Figure \ref{fig2}.b). This second point is crucial to realize complex circuits such as SQUIDs (Superconducting QUantum Interference Devices) \cite{cleuziou2006carbon, schneider2012coupling}, or charge detectors \cite{waissman2014carbon}, which require accurate positioning of the CNT. The metal used to contact the CNT is essential to obtain a low resistance CNT-metal interface. Palladium has been shown to be an excellent candidate \cite{javey2003ballistic} mainly because of its low workfunction mismatch with CNTs. We thus use palladium as a top layer.

\begin{figure*}[ht]
\centering
\includegraphics[width=1\linewidth]{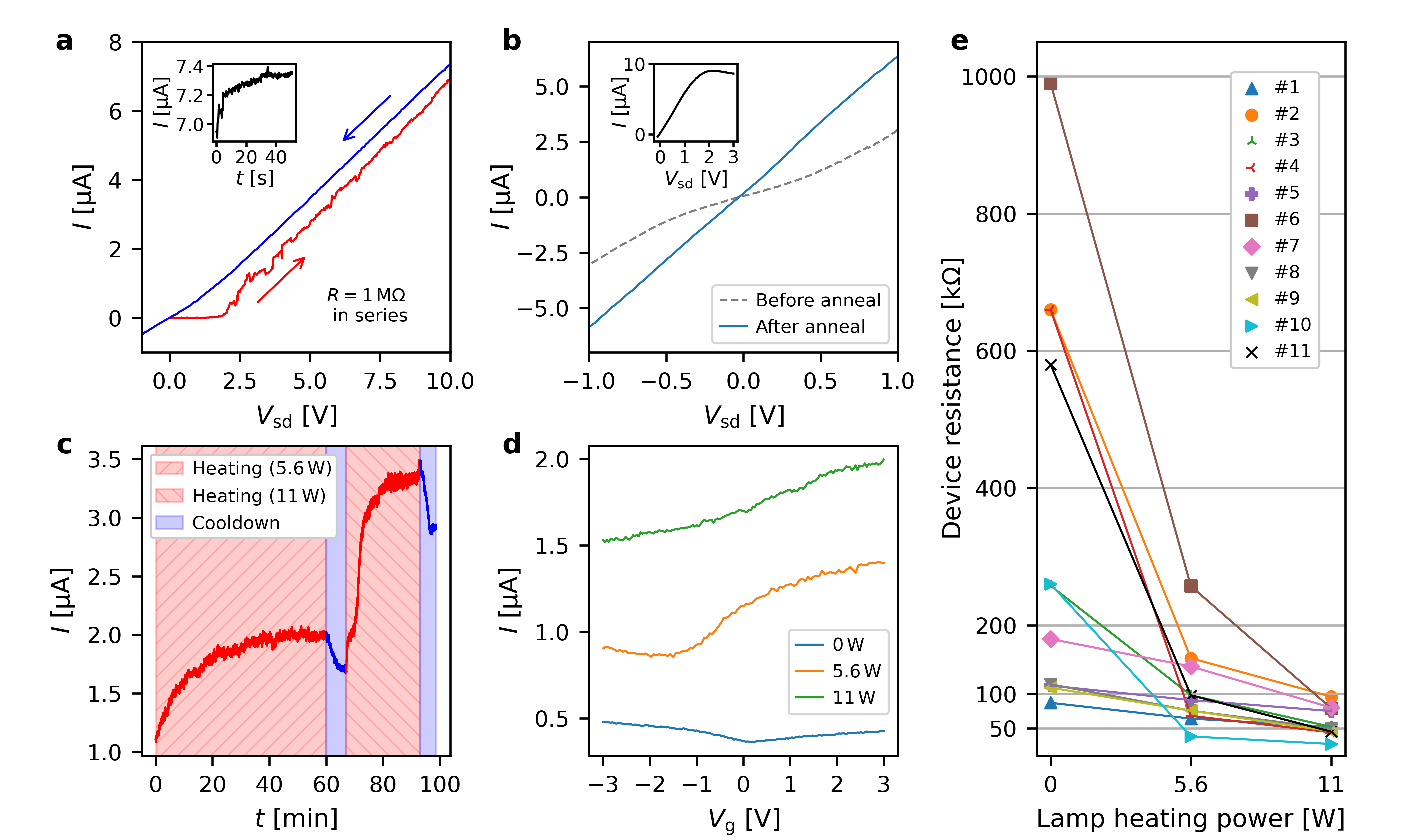}
\caption{\textbf{Two steps annealing technique.}
\textbf{a.} Trace-retrace cycle of an $I$-$V$ curve during the first current-induced annealing step with 1$\,$M$\Omega$ in series. The inset shows the time evolution of the current between the trace and the retrace.
\textbf{b.} $I$-$V$ curves measured before and after the second current-induced annealing step. The current measured during this annealing step is shown in the inset.
\textbf{c.} Current measured in an ancillary CNT junction during the two steps of RTA at a bias of $0.1\,$V.
\textbf{d.} Current as a function of gate voltage measured on device $\#$10 at different stages of the process. The "0$\,$W" label corresponds to the moment just after the current-induced annealing was finished. Curves labeled "5.6$\,$W" and "11$\,$W" are measured after the RTA step at the corresponding power.
\textbf{e.} Evolution of device resistances of 11 transferred CNTs during the full annealing process (measured at a bias of $0.1\,$V).}
\label{fig4}
\end{figure*}

In order to obtain a high quality CNT-metal interface, we paid great attention to the cleanliness of the circuit as well as the CNT throughout the full assembly process. First of all, we remove adsorbates from the surface of the circuit before the transfer of the CNT using an argon-milling process calibrated to etch 5$\,$nm of palladium (see Figure \ref{fig1}.b). Indeed, the first 3-4$\,$nm of palladium can adsorb oxygen \cite{jung2021understanding} which is known to deteriorate the quality of the CNT-metal interface \cite{felten2009role, jung2021understanding, collins2000extreme}. The resistance improvement due to this pre-cleaning step is quantified in Figure \ref{fig3}. In addition, to maintain a clean surface of the CNT, we developed a transfer technique allowing to maintain the CNT in vacuum from their synthesis to their measurement in a cryostat ($P < 5 \times 10^{-2} \,$mbar during the transfer steps). We believe this strongly reduces the amount of adsorbates at its surface. Finally, because the exposure to the e-beam can induce the deposition of hydrocarbons (see Supplementary Note 2), the region of the circuit where the CNT is transferred onto, is never exposed to the e-beam. For the same reason, the section of the CNT which will be connected is only minimally exposed during localization of the CNT (see Methods). 

\begin{figure*}[ht]
\centering
\includegraphics[width=0.85\linewidth]{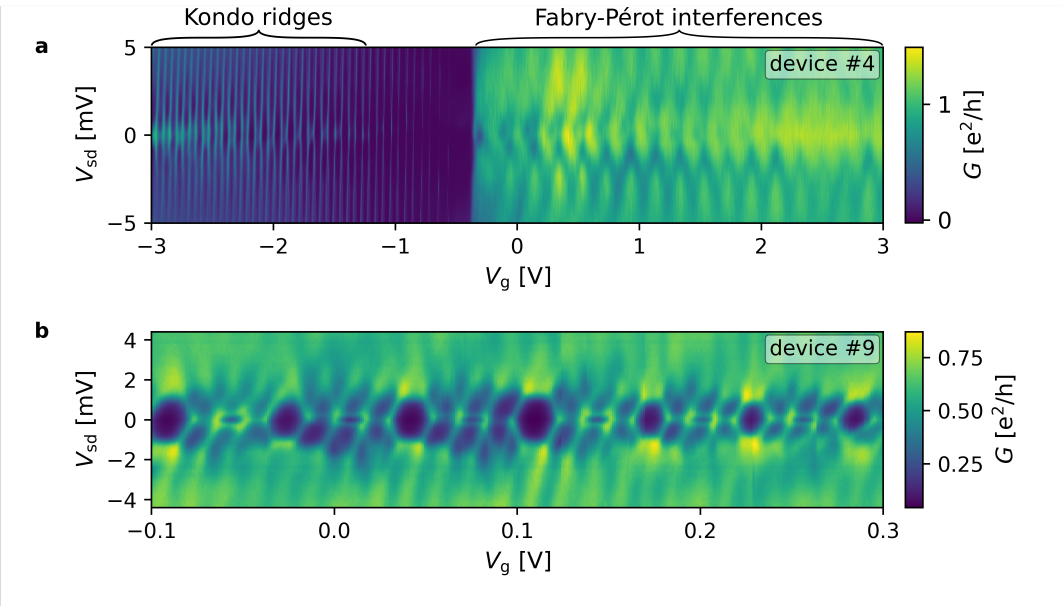}
\caption{\textbf{Kondo regime and Fabry-Pérot interferences.} 
Maps of the differential conductance measured at 30$\,$mK as a function of the source-drain bias $V_{\mathrm{sd}}$ and gate voltage $V_{\mathrm{g}}$. \textbf{a.} device $\#$4, \textbf{b.} device $\#$9.}
\label{fig5}
\end{figure*}

\subsubsection*{Current-induced and radiative thermal annealing steps}
Nevertheless, having a clean CNT-metal interface is not sufficient to obtain a highly transparent contact, hence we developed a two-step low temperature (below 300$\,$°C) annealing process in which the contact is first improved by driving a large current through the CNT (see Figure \ref{fig1}.a and \ref{fig1}.b) and then further enhanced via RTA (see Figure \ref{fig1}.d and \ref{fig1}.e). 

The first current-induced annealing technique is a very local process since most of the heating power is dissipated at the contact between the CNT and the metal (see Figure \ref{fig1}.d). $I$-$V$ curves during current-induced annealing are presented in figures \ref{fig4}.a, and \ref{fig4}.b. In order to prevent an accidental cut of the CNT, a first annealing is performed with a $1\,\mathrm{M}\Omega$ resistor in series. On the voltage ramp-up, a current plateau is visible at low bias voltage which is reminiscent of a barrier forming at the CNT-metal interface. It is followed by sudden current jumps suggesting discrete events in the vicinity of the CNT-metal interface, such as atomic rearrangements, or desorption of atoms, which are consistent with the local heating scenario. After this first trace-retrace cycle, the initial current plateau has disappeared, which confirms the decrease of the barrier due to the local events. The CNT junction still displays a weak non-ohmic behavior, which is eliminated by driving the maximum current that the CNT can sustain \cite{pop2005negative}.

The second annealing step consists in a RTA of the entire surface of the device. For this purpose, we use the thermal radiation of a 70W halogen lamp which has been installed inside the SEM vacuum chamber. During the RTA step the device is positioned about 5$\,$mm below the lamp. A tantalum shield partially surrounds the halogen lamp to prevent excessive heating of the SEM column. The RTA is subdivided in two steps, at 5.6$\,$W and 11$\,$W, both ranging from 30 to 60 minutes for most of the devices (see Supplementary Note 6). These power values gave the best result for the different metals used for the bulk of the contact electrodes (Ti, Mo, Nb, see Supplementary Note 6). In contrast to the current-induced annealing, the RTA displays a more continuous evolution of the device resistance (see Figure \ref{fig4}.c and Supplementary Note 4). During the cooldown periods, the resistance increases again which we associate to some thermoelectric effect occurring at metal junctions in the circuit.

The effect of the different annealing steps on the resistance of 11 devices is presented in Figure \ref{fig4}.e. Importantly, one can notice that all devices have a resistance above 50$\,$k$\Omega$ after the current-induced annealing, while 7 out of the 11 devices are below this value after the full annealing procedure, showing the substantial effect of RTA. Device $\#$10  could reach a minimal resistance of 25$\,$k$\Omega$. Besides, a resistance of 300$\,$k$\Omega$ after the current-induced annealing step strongly helps to reach a final resistance inferior to 50$\,$k$\Omega$. It shows that an initially clean interface is required to reach final low resistance values, underlining the need for an argon-milling pre-treatment step. Moreover, the number of devices for which this behavior has been observed further emphasizes the robustness of our technique.

Another effect of the RTA is to switch the CNT transport behavior from ambipolar to unipolar n-type (see Figure \ref{fig4}.d). This change is accompanied by a shift of the center of the semiconducting gap towards negative gate values. It can be explained either by the modification of the workfunction mismatch between the palladium and the CNT, or by the doping of the CNT \cite{derycke2001carbon, derycke2002controlling}. While the exact mechanism taking place in our situation is not clear, both can result from the removal of adsorbates either at the CNT-metal interface or on the CNT itself \cite{chen2006tuning}. In particular, oxygen desorption has been shown to have a dominant effect in similar circuits \cite{derycke2001carbon, derycke2002controlling}. Interestingly, the n-type behavior resulting from our approach is in clear opposition to the more common p-type behavior of most CNT-based open quantum circuits realized so far.

It is important to note, that this annealing procedure only requires relatively low temperatures estimated to not exceed 300$\,$°C (see Supplementary Note 3). This temperature has to be compared to the range of 800-900$\,$°C involved in the CNT top-growth fabrication technique \cite{cao2005electron}, as well as the 800$\,$°C needed to form CNT-metal covalent bonds (MoC \cite{cao2015end}, or TiC formations \cite{martel2001ambipolar, zhang1999heterostructures}). We believe RTA is particularly efficient since it directly heats the surface of the circuit which is where the contact between the CNT and the metal is located. This relatively low temperature annealing technique is thus compatible with other quantum technologies such as superconducting circuits (based on Al, Nb, NbTi, ...).

\subsubsection*{Open quantum dot regime}
The devices with a resistance below 50$\,$k$\Omega$ were further studied with cryogenic transport measurements (see Figure \ref{fig5} and Supplementary Note 5). In both charge stability diagrams presented in Figure \ref{fig5}, the presence of Kondo-ridges demonstrates that cotunneling processes play a dominant role in the transport spectrum of the device, proving the presence of large tunnel coupling rate between the quantum dot and the Fermi reservoirs \cite{nygaard2000kondo}, but still lower than the charging energy of the quantum dot ($k_B T < \Gamma < E_C$). Even higher coupling could be reached as shown by the Fabry-Pérot interference pattern in Figure \ref{fig5}. In this regime, the tunnel coupling rates to the Fermi reservoirs are larger than the charging energy of the quantum dot ($k_B T < E_C < \Gamma$), therefore the electron number in the quantum dot is not a good quantum number anymore, proving that it cannot be considered as a closed quantum system. Moreover, the observation of the four-fold degeneracy of the CNT energy levels confirms the very low amount of disorder in the circuit.

\subsection*{Conclusion}

To conclude, we demonstrated that the stamping nano-fabrication technique is able to produce CNT-based quantum circuits deep into the open quantum dot regime. For this, we developed a technique relying on the combination of an argon-milling pre-treatment, vacuum transferred CNTs and a two-step annealing procedure, enabling us to reach device resistance values below 50$\,$k$\Omega$. Such low resistance values are sufficient to reach the open quantum dot regime in which cotunneling processes dominate, leading to Kondo physics as well as Fabry-Pérot interferences. Moreover, our technique allows to integrate a CNT into a circuit with a precision of $\pm\,200\,$nm. A more meticulous approach combined with dedicated markers should in principle allow to reach a precision down to $\pm\,50\,$nm.

The accuracy of this technique, the relatively low required temperature (below 300$\,$°C) and its capability to reach the open quantum regime make it particularly attractive to realize hybrid systems and merge different mature quantum technologies. For instance, one can use a suspended CNT connected with superconducting electrodes to realize a Josephson junction which could be used as the basic element of novel superconducting qubits. The low Josephson energy obtained in nanotube \cite{mergenthaler2021circuit} could serve to explore new areas of the parameter space of superconducting qubit \cite{pechenezhskiy2020superconducting}.

The developed technique is also perfectly suited to realize optomechanics experiments based on CNTs. One promising approach relies on the recently studied flux-mediated phonon-photon coupling \cite{rodrigues2019coupling, schmidt2020sideband}, for which the large zero-point fluctuations of the CNT are a powerful advantage \cite{khosla2018displacemon}. Using such a coupling scheme, a single-phonon coupling rate g$=1\,$MHz is within reach (see Supplementary Note 7). This value is sufficient to reach the single photon strong coupling regime allowing to create mechanical quantum state in the CNT such as Schrödinger cat states \cite{hauer2022nonlinear}.

\subsection*{Methods}
\paragraph{CNT growth}
The CNTs are grown using a chemical vapor deposition (CVD) process on a custom cantilever comb chip (see Supplementary Note 1) which allows us to transfer CNTs suspended between cantilevers onto the circuit chips. The growth is performed at 880$\,^\circ$C for 15 min with precursor gas flows of 75$\,$sccm of CH$_4$ and 50$\,$sccm of H$_2$. We use the same catalyst as in ref \cite{viennot2014stamping}. 
\paragraph{Circuit chip fabrication}
The fabrication of the circuit chip comprises four layers of lithography processes on a p-doped SiO$_2$ (300$\,$nm) / Si (500$\,$µm) substrate. First, markers and large structures are patterned with optical lithography and electron-beam evaporation of Ti(3$\,$nm) / Au(40$\,$nm). Gate (Ti/Al/AlOx) and contact electrodes (Ti/electrode material/Pd) are then patterned with electron-beam lithography and electron-beam evaporation with a lift-off process.

In order to transfer a CNT from the cantilever onto the circuit chip, 12$\,$µm deep trenches are patterned with optical lithography and reactive-ion etching using a combination of CHF$_{3}$/Ar/O$_{2}$ (SiO$_2$ etching) and SF$_{6}$/C$_{4}$F$_{8}$ (Si etching).

\paragraph{Integration of the CNT}
We transfer the cantilever chip from the CVD oven into an SEM with a custom holder using a vacuum transfer arm. The top 60$\,$µm of the cantilever chip are then imaged to preselect the suspended CNT to be stamped (panorama image presented in Supplementary Figure S1). Prior to stamping, the surface of the circuit chip is cleaned with an argon-milling process using a Kaufman source (for 90$\,$s, at a discharge voltage of 40$\,$V, with a flow of 9$\,$sccm Ar and a beam acceleration voltage of 200$\,$V). This removes about 5$\,$nm of Pd including the oxygen adsorbed at the surface \cite{jung2021understanding}. 

The circuit chip is then micro-bonded and transferred into the SEM usually within 15-20$\,$min. Because oxygen diffusion at the Pd surface takes place over several hours \cite{jung2021understanding}, we believe the short exposure time to air of 15-20 min after argon-milling is not detrimental to the quality of the surface. The cantilever pair with the preselected suspended CNT is laterally aligned with the stamping position on the sample chip using piezomotors (minimum step size: 10$\,$nm). The cantilevers are then lowered into the trenches of the circuit chip while tracking the distance between the CNT and the surface of the circuit with the SEM. Electron-beam exposure of the whole tube is limited to the quick overview panorama image (see Supplementary Note 1). During the approach only few microns of the tube edge are exposed to the e-beam while the central part remains unexposed. We apply a bias of 4$\,$V between the two outermost electrodes and the next inner electrodes and once the CNT is in mechanical contact a current through it is measured showing that it has been contacted electrically. To retract the cantilever chip without affecting the CNT segment of interest, the circuit chip contains dedicated cutting sections where the CNT junction can be cut by driving a high current ($\gtrsim\,$10$\,$µA) through the CNT. After annealing the individual junctions by driving a high current through the CNT, the chip is radiatively heated with the halogen lamp. Circuit chips with CNT junctions of resistance below 50$\,$k$\Omega$ at a bias of 0.1 V are then vacuum-transferred to a dilution cryostat with 30$\,$mK base temperature where the low-temperature measurements are performed. During the whole process, the CNTs are thus never exposed to air.

\subsection*{Acknowledgement}
We thank the KIT Nanostructure Service Laboratory (NSL) for the technical support in the nanofabrication of the devices. T.A. acknowledges financial support from the Hector Fellow Academy. All authors acknowledge support from the Leibniz award WE 4458-5, and the DFG grant WE 4458/2-1.

\subsection*{Data availability}
The data that support the findings of this study are available from the corresponding author upon reasonable request.

\subsection*{Author contributions}

T.A, T.C, and A.A performed the measurements. T.A and T.C fabricated the devices with the help of A.A. T.A performed the data analysis with the assistance of T.C and A.A. T.A and T.C developed the nano-assembly method with the help of A.A, C.S, and W.W. T.C developed the experimental setup with the help of T.A, A.A, C.S, and W.W. T.C wrote the manuscript, and T.A wrote the supplementary materials. A.A and W.W proofread the manuscript and the supplementary materials. 

\subsection*{Competing interests}
The authors declare no competing interests.

\printbibliography

\end{document}


\maketitle
\tableofcontents

\section{Supplementary Note 1. CNT growth chips}
\label{SI_note1}
\begin{figure}[htbp]  
	\centering
	\includegraphics[width=0.85\textwidth]{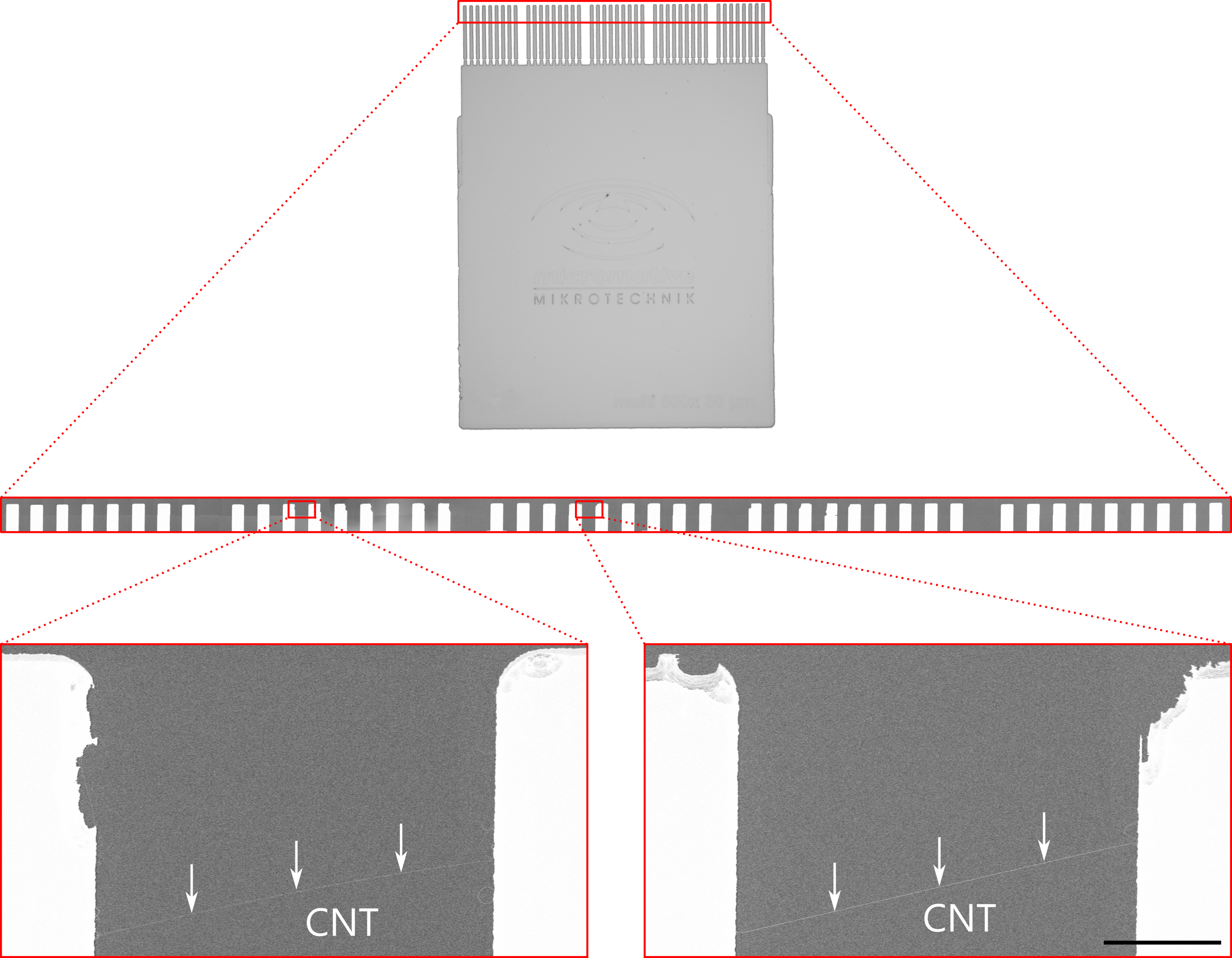}
	\caption{Overview of a cantilever chip with the panorama scanning electron microscope (SEM) image taken to locate the suspended CNTs (marked by arrows). The left CNT has been integrated into device \#4. The chip measures $3\,$mm in width and the scale bar measures $10\,$µm.}\label{panorama_canti55} 
	
\end{figure}

The CNTs are grown on custom cantilever comb chips (cf. Supplementary Figure \ref{panorama_canti55}) which allows us to transfer CNTs suspended between cantilevers onto the circuit chips. An overview of such a cantilever chip is shown in Supplementary Figure \ref{panorama_canti55}.

\section{Supplementary Note 2. E-beam induced hydrocarbon deposition minimization}
\label{SI_note2}

The SEM imaging of the carbon nanotubes during the localization step is expected to lead to e-beam induced hydrocarbons (HC) deposition on the nanotube \cite{hugenschmidt2023electron}. The presence of HC at the surface of the carbon nanotube can deteriorate the quality of the metal-nanotube interface and possibly lead to charge fluctuators in the direct vicinity of the carbon nanotube quantum dot. 

Nevertheless in our situation we expect the HC deposition on the suspended nanotubes to be very low, and this for several reasons. It was shown that HC deposition is mainly due to the low energy ($<50\,$eV) secondary electrons (SE) \cite{burbridge2008strategies} which result from the interaction of the primary electrons with atoms. Having a low number of SEs in the vicinity of the nanotube is thus a key point to minimize the HC deposition at its surface. In our situation, the section of the nanotube which is transferred into the circuit corresponds to the center of a 30µm long suspended nanotube. This means that the amount of atoms in the vicinity of the transferred section of the nanotube is extremely low. Consequently the production of SEs close to the transferred section of the nanotube is expected to be much lower than the one in the neighborhood of a nanotube lying on a substrate.

In addition, we use a relatively high acceleration voltage ($8\,$kV), which results in a lower number of SE out of the interaction with the nanotube in comparison to a low acceleration voltage beam ($1-2\,$kV) \cite{hugenschmidt2023electron}. In Supplementary Figure \ref{HC}, we show that the acceleration voltage of the e-beam has a strong influence on the amount of HC deposited at the surface of a silicon substrate. One visible consequence is that the contrast between the suspended nanotube and the background is low, while the electron count on the cantilevers is completely saturated (white on SEM picture in Supplementary Figure \ref{panorama_canti55}). This indicates the low amount of SEs produced in the vicinity of the nanotube.

A second important factor is the amount of precursors species for HC formation, mainly coming from the residual vacuum and the surfaces in the chamber \cite{hugenschmidt2023electron}. We thus work at low pressure ($P= 2.5 \times 10^{-6}\,$mbar) and because the nanotube is suspended and is consequently far away from surfaces, one can expect a lower HC deposition on the nanotube.

The parameters used during the exposure are:

\begin{itemize}
    \item Acceleration voltage: $8\,$kV
    \item Chamber pressure: $P= 2.5 \times 10^{-6}\,$mbar
    \item Beam current: $70\,$pA
    \item Pixel size: $27.9\,$nm $\times$ $27.9\,$nm
    \item Exposure time per pixel: $2.5\, \mu$m
\end{itemize}

\begin{figure}[htbp]  
	\centering
	\includegraphics[width=1\textwidth]{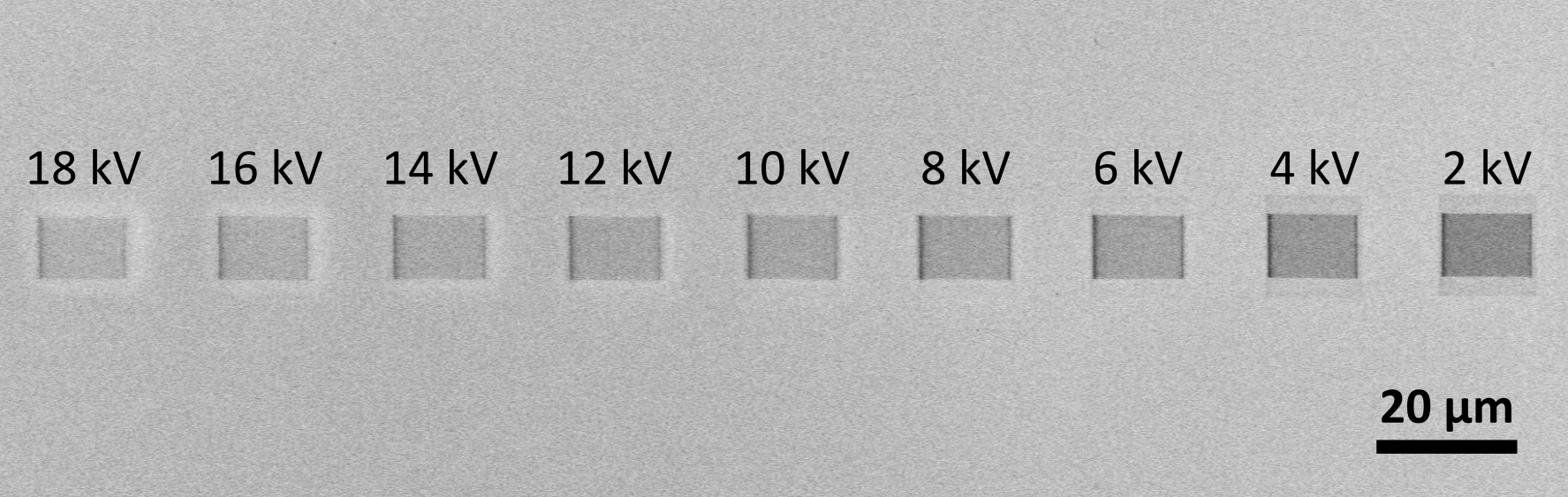}
	\caption{The lower amount of HC deposition the higher is the acceleration voltage is due to the fact that secondary electrons are produced deeper under the surface of the substrate at high acceleration voltage.}\label{HC} 
	
\end{figure}

\section{Supplementary Note 3. Radiative thermal annealing (RTA) - process and temperature estimation}
\label{SI_note3}

\begin{figure}[htbp]
	\centering
	\includegraphics[width=0.35\textwidth]{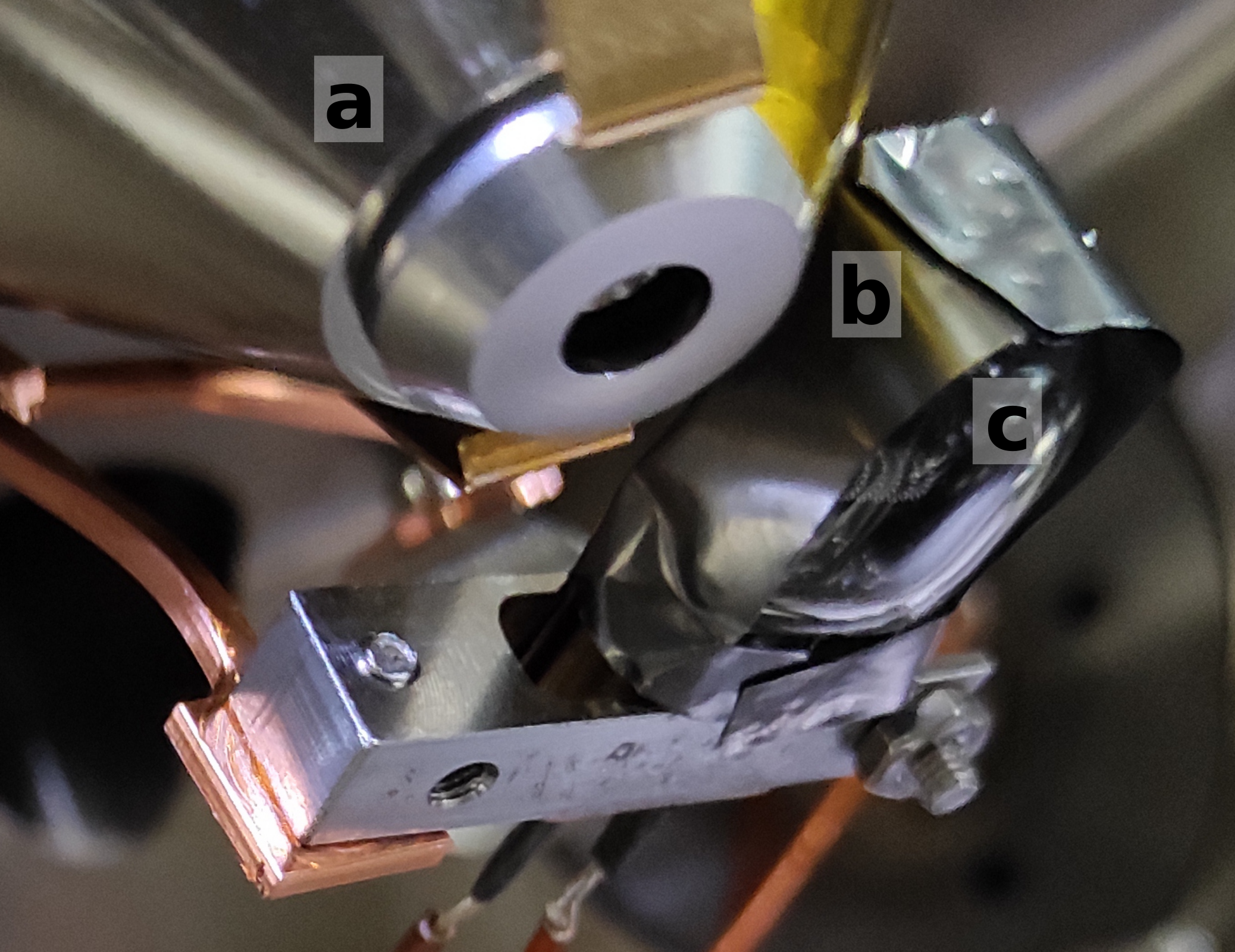}
	\caption{Image of the radiative thermal annealing setup inside the SEM with \textbf{a.} e-beam column, \textbf{b.} Ta shield and \textbf{c.} halogen lamp with filament.}
	\label{lamp}
\end{figure}
\begin{figure}[htbp]
	\centering
	\includegraphics[width=1\textwidth]{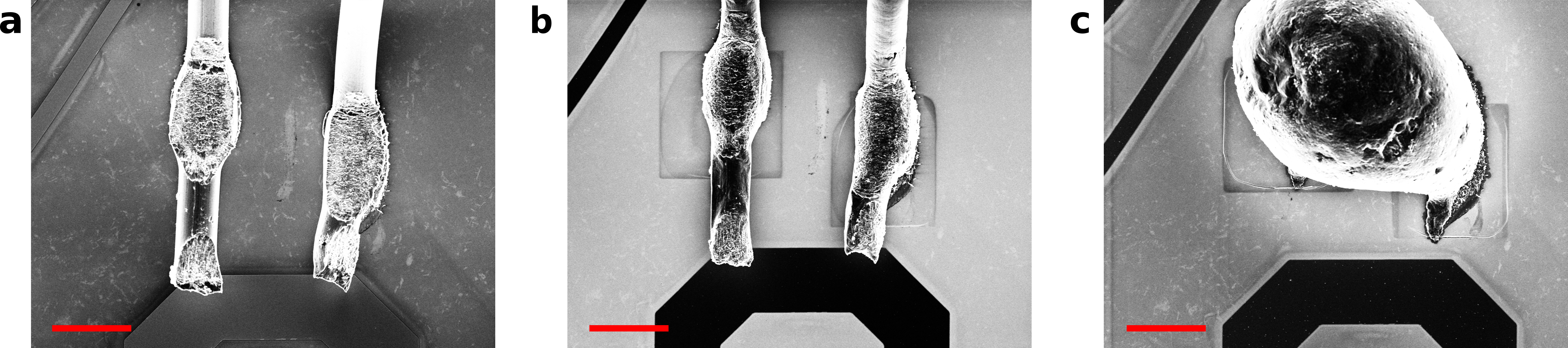}
	\caption{SEM image of a circuit chip with aluminium bond wires. \textbf{a.} Before RTA, \textbf{b.} After RTA at $26.2\,$W, \textbf{c.} After RTA at 29.5$\,$W. Scale bar: $50\,$µm.}
	\label{Almelting} 
\end{figure}
After current-induced annealing of the individual CNT junctions, the lamp is preheated to remove potential adsorbates while monitoring the pressure inside the vacuum chamber of the SEM (degasing at 16$\,$W, 5$\,$min). For this purpose, the chip is moved far away from the e-beam column and the halogen lamp (cf. Supplementary Figure \ref{lamp}). The circuit is then placed about $5\,$mm below the lamp and the chip is radiatively heated at a lamp power of 5.6$\,$W for a time varying from 25 to 100$\,$min. A second heating step is performed at 11$\,$W. The heating time for this step has been varied between 5$\,$min and 45$\,$min. The CNT junctions are characterized after each step once the chip has cooled down again.

Measuring the current through the CNT during RTA allows us to monitor the physical processes at the CNT-metal interface and to find the optimum duration which varies for each sample. In general, there is only little contact improvement below $5\,$min. For RTA beyond 30-45$\,$min the measured current remains nearly constant for all devices independent of both powers. We therefore consider 30-45$\,$min to be the best RTA duration. A further heating at 16$\,$W did not show any contact improvement and led to higher junction resistance in some cases.
\\\\
For a better understanding of the RTA process it is essential to know the temperature at the surface of the circuit chip. Performing a standard temperature measurement using a thermocouple would be inaccurate because the thermal dissipation through the device substrate and the sample holder is radically different from the dissipation via the thermocouple wires. Therefore, in order to estimate the temperature at the surface of the circuit chip during RTA we relied on the melting temperature of aluminium bond wires. We micro-bonded aluminium bond wires across the chip and performed heating tests at different lamp powers inside the SEM at a pressure in the low $10^{-5}\,$mbar range and for $30\,$min each (cf. Supplementary Figure \ref{Almelting}). At a power of $26.2\,$W the bonds still remained intact and only the surface is partially molten while they are completely molten at 29.5$\,$W. Putting these values in reference with the melting temperature of bulk aluminium of $660^\circ$C and assuming a linear relationship between lamp power and chip temperature we can roughly estimate the sample temperature to 200-300$\,^\circ$C at $11\,$W and 100-150$\,^\circ$C at $5.6\,$W. This temperature range is much smaller than the typical growth temperature of 800-900$\,^\circ$C for CNTs grown on top of metal electrodes. 
Because aluminium wire bonds are much larger than the typical dimensions of the circuit (30$\,$µm width compared to about 1$\,$µm for the circuit electrodes), and aluminium has one of the largest thermal conductance among the used metals, we expect the thermal dissipation to be stronger in the aluminium wire bonds than at the surface of the circuit electrodes. For this reason, our method to estimate the temperature only provides lower bounds of the real temperature at the surface of the circuit. Moreover, the exact surface temperature strongly depends on the used metal.

\section{Supplementary Note 4. Further examples for gate dependence measurements after current-induced annealing and RTA}
\label{SI_note4}
We measured the gate dependence after current-induced annealing and after different radiative thermal annealing steps for more devices as well. All devices with a narrow gap in the gate dependence showed a behavior similar to the device shown in figure 4 in the main part.
\begin{figure}[hbtp]
	\centering
\includegraphics[scale=0.9]{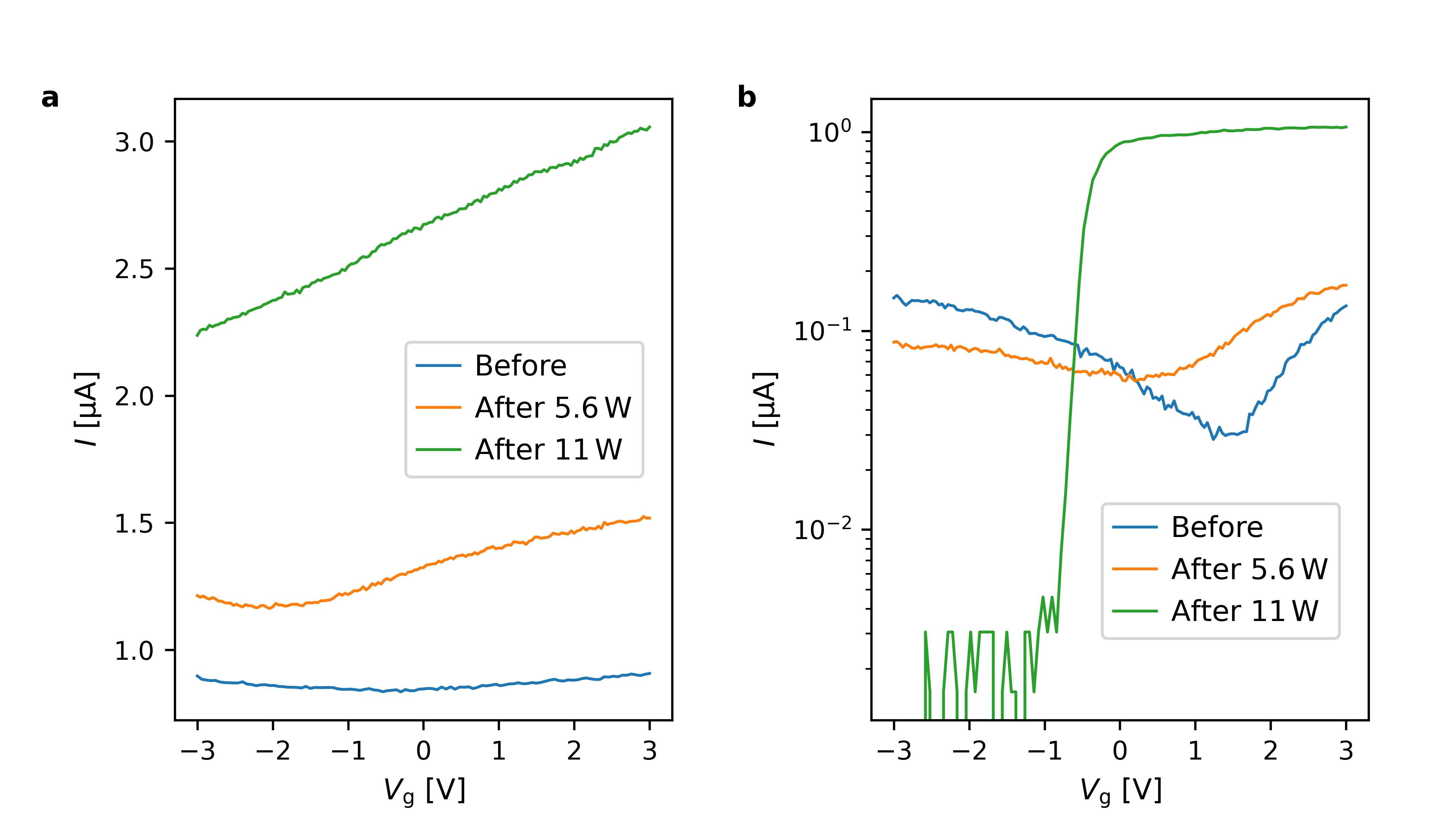}
	\caption{Gate dependence of two other devices at a source-drain voltage $V_\text{sd}=0.1\,$V before the RTA and after the two RTA steps. \textbf{a.} Device \#9. \textbf{b.} Device \#12. This device has only been measured at higher voltages than $V_\text{sd}=0.1\,$V before the second RTA step and the current has been scaled to $V_\text{sd}=0.1\,$V, assuming ohmic behavior.}
\end{figure}

\section{Supplementary Note 5. Additional low temperature measurements}
\label{SI_note5}
In this section we present further cryogenic measurements of different devices. Unless specified otherwise the measurements were carried out at the base temperature of the cryostat (30$\,$mK). These cryogenic measurements complement the ones from figure 5 in the main text and emphasize that we can reach highly transparent contacts between the CNT quantum dot and the leads in multiple devices (cf. Fabry-Pérot resonances (Supplementary Figure \ref{map_appendix_FPR}) and Kondo ridges (Supplementary Figure \ref{map_appendix_Kondo})), while the observation of fourfold shell filling in Supplementary Figure \ref{vgvg} demonstrates the cleanliness of our devices. In these measurements the differential conductance value is measured for the whole chip and includes the resistance of the circuit electrodes. The exact line resistances on chip vary with the electrode material but are usually in the range of few k$\Omega$.

\begin{figure}[hbtp]
\centering
\includegraphics[scale=0.94]{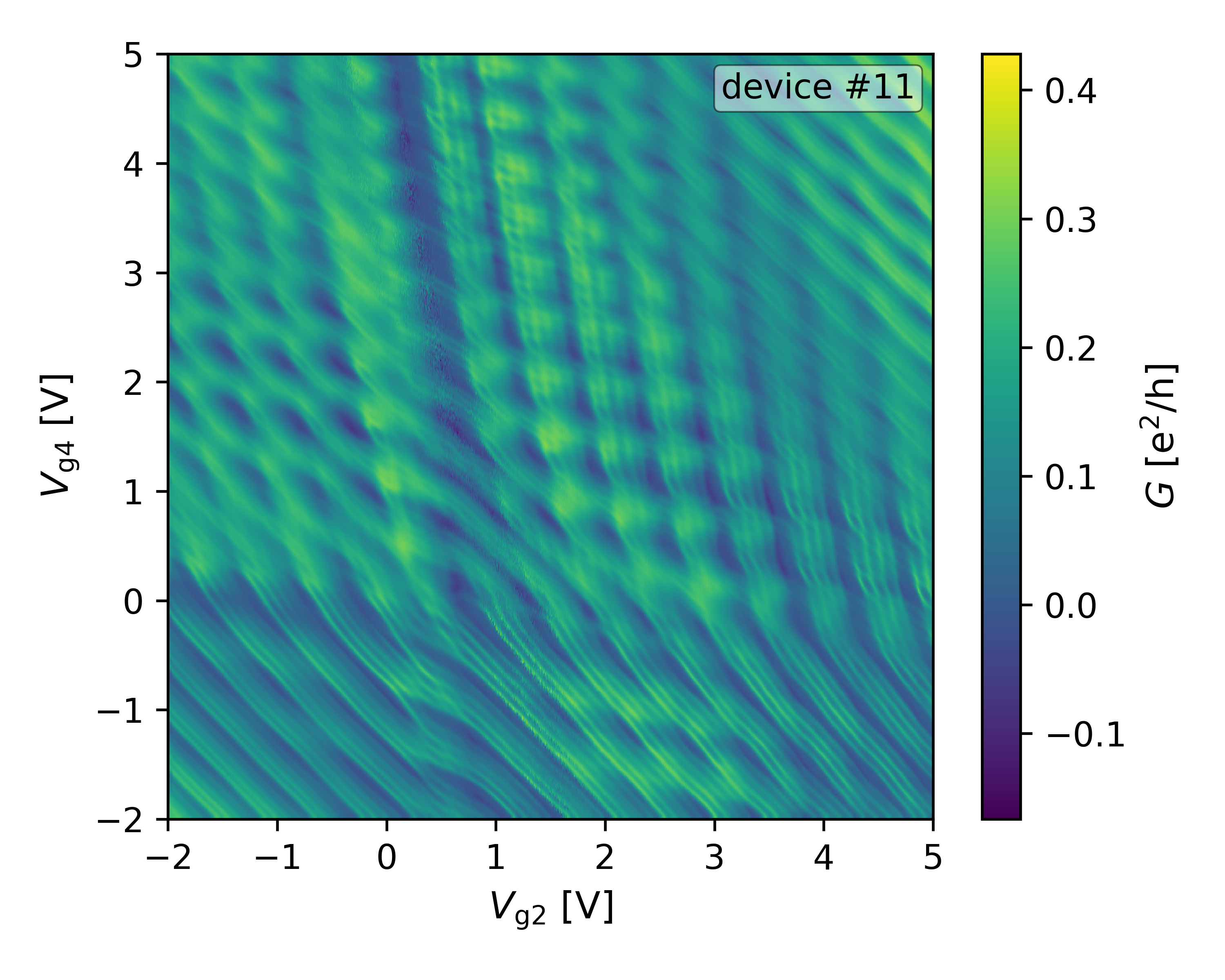}
\caption{Double quantum dot charge stability diagram at source-drain voltage $V_\text{sd}\approx 0\,$V for device \#11 showing clear four fold shell filling. The second and fourth gate out of five gates below the suspended CNT are varied while $V_\text{g1}=V_\text{g3}=-2\,$V and $V_\text{g5}$ is floating. The device is similar to the one presented in figure 2a in the main text. The measurement is performed at $30\,$mK.}
\label{vgvg}
\end{figure}

\begin{figure}[hbtp]
	\centering
	\includegraphics[scale=0.94]{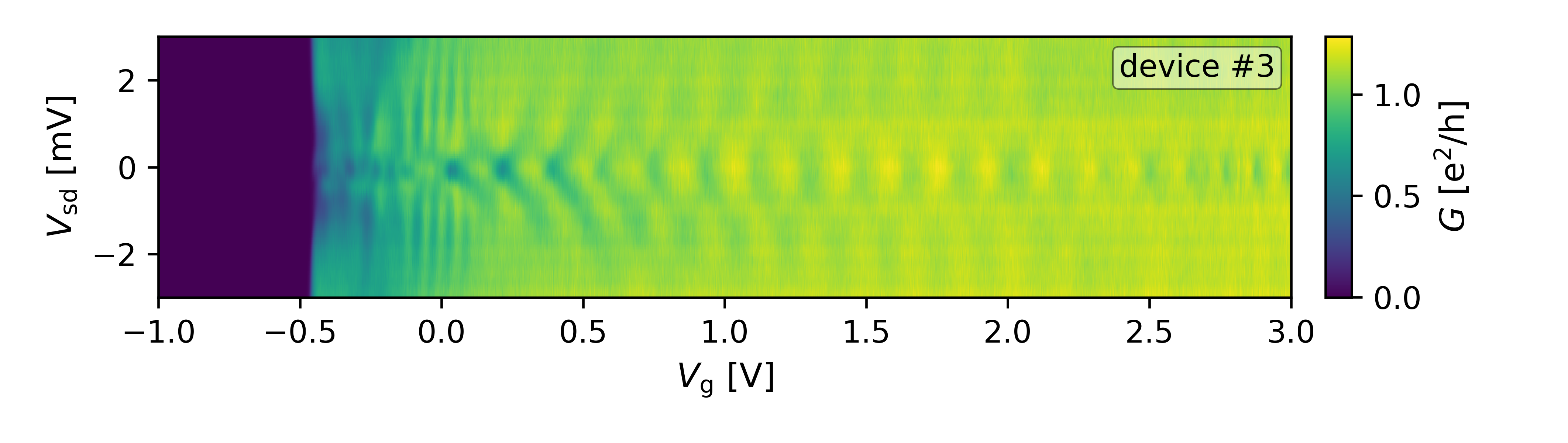}
	\caption{Source-drain bias $V_\text{sd}$ vs. gate voltage $V_\text{g}$ conductance map of device \#3 with Fabry-Pérot resonances for positive gate voltage, measured at $150\,$mK.}\label{map_appendix_FPR}
\end{figure}

\begin{figure}[hbtp]
	\centering
	\includegraphics[scale=0.94]{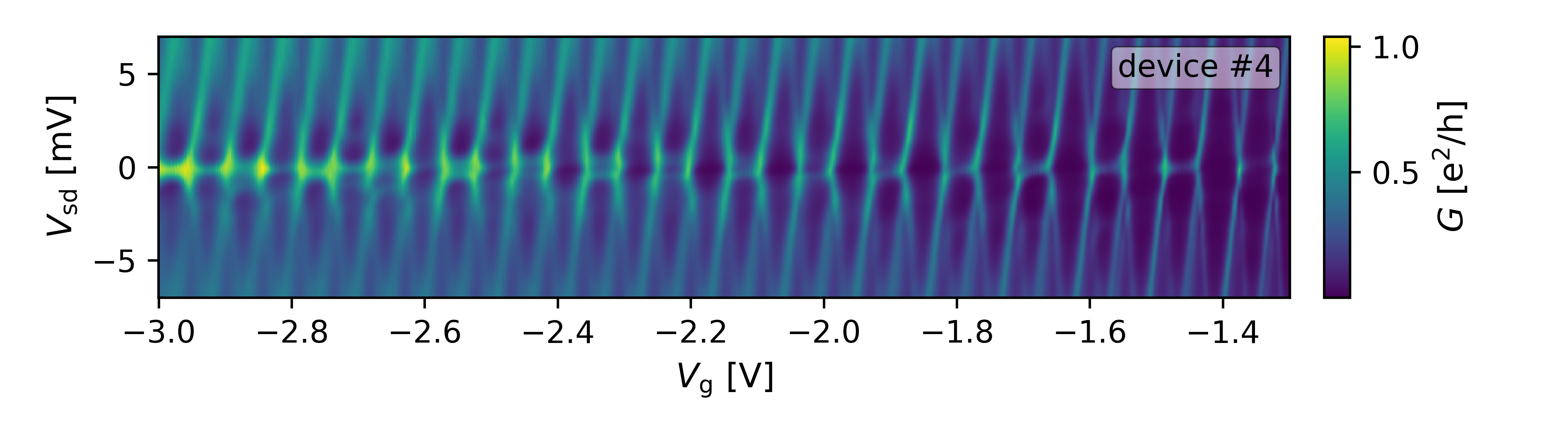}
	\caption{Source-drain bias $V_\text{sd}$ vs. gate voltage $V_\text{g}$ conductance map with a zoom into the negative gate voltage range of the map shown in figure 5a in the main text (device \#4). Kondo ridges are visible between every second diamond and towards -3$\,$V we can observe a slight transition from SU(2) to SU(4) Kondo effect. The measurement is performed at $30\,$mK.}\label{map_appendix_Kondo}
\end{figure}

\section{Supplementary Note 6. Device details}
\label{SI_note6}
In Supplementary Table \ref{tab:data} we provide details about the devices presented in the main text and supplementary notes. Device \#12 has been excluded in figure 4e because it has not been measured at $V_\text{sd}=0.1\,$V before heating.

 \begin{table}[ht]
	\centering		
	\caption{Overview of the devices presented in figure 4 in the main text and in the supplementary notes. $t_\text{Ar} $ denotes the time of argon milling, $t_\text{air}$ the time of air exposure between argon milling and loading of the sample in the SEM, $t_\text{heat1}$ the radiative thermal annealing time at $5.6\,$W and $t_\text{heat2}$ the time at 11$\,$W.}
\begin{tabular}{cccccc}
	\toprule
device 	& electrode material& $t_\text{Ar} $ [min] & $t_\text{air}$ [min] & $t_\text{heat1}$ [min] & $t_\text{heat2}$  [min]\\
		\midrule
\#1	& Ti (5nm), Nb (90nm), Pd (8nm) & 1.5
 & 34 & 30 & 12.5 \\
\#2	& Ti (5nm), Nb (90nm), Pd (8nm) &1.5
& 34 & 30 & 12.5 \\
\#3	& Ti (5nm), Nb (90nm), Pd (8nm) &1.5& 32 &28  & 13 \\

\#4	& Ti (5nm), Nb (90nm), Pd (8nm) &1.5& 43 & 30 & 5.5 \\

\#5 	& Ti (5nm), Nb (90nm), Pd (8nm) &2 & 27 & 50  & 30 \\

\#6	& Ti (5nm), Nb (90nm), Pd (8nm) &2&  27 & 50 & 30 \\
	
\#7	& Ti (90nm), Mo (8nm), Pd (5.5nm) &1.5& 32 & 60  & 25 \\
	
\#8	& Ti (90nm), Mo (8nm), Pd (5.5nm) &1.5&	14  & 100 & 30 \\
	
\#9	& Ti (90nm), Mo (8nm), Pd (5.5nm) &1.5& 13 & 40 & 30 \\
	
\#10	& Ti (90nm), Pd (5.5nm)&1.5 & 14 & 45 &  45\\
	
\#11	& Ti (90nm), Pd (5.5nm) &1.5& 14 & 45 & 45 \\

\#12	& Ti (90nm), Pd (5.5nm) &1.5& 20 & 30 & 30 \\
		\bottomrule
\end{tabular}
\label{tab:data}
\end{table}

\section{Supplementary Note 7. Flux mediated single photon coupling estimation}
\label{SI_note7}
We consider a circuit where a SQUID (Superconducting Quantum Interference Device) built out of a carbon nanotube \cite{cleuziou2006carbon, schneider2012coupling} is integrated in a microwave cavity similarly to what has been realized with aluminium suspended beams \cite{rodrigues2019coupling, schmidt2020sideband}. In such systems, the photon energy depends on the mechanical oscillations of the carbon nanotube via the modulation of the magnetic flux threading the SQUID loop. The flux-dependent resonance frequency of the cavity reads: 
\begin{equation} \label{res_freq}
\omega(\phi_S) = \frac{1}{\sqrt{(L + L_{S}(\phi_S))C}} = \frac{1}{\sqrt{(L + \frac{\phi_0}{2 I_{C,JJ} \lvert cos(\pi \phi_S / \phi_0) \rvert})C}}
\end{equation}
Where $L$ and $C$ are the inductance and capacitance of the bare cavity respectively, $L_S$ is the inductance of the SQUID, $\phi_S = \Phi_S / 2 \pi$ is the reduced magnetic flux threading the SQUID loop, $\phi_0 = \Phi_0 / 2 \pi = \hbar / 2e $ is the reduced magnetic flux quantum, and $I_{C,JJ}$ is the critical current of a single Josephson junction. Here we neglected the linear inductance of the arms of the SQUID. The single-photon, single-phonon optomechanical coupling rate resulting from this interaction is:

\begin{equation} \label{g0_exp}
g_0 = \frac{\partial \omega_0}{\partial \phi} \phi_{ZPF} = \frac{\partial \omega_0}{\partial \phi} \gamma B_{\parallel} l x_{ZPF}
\end{equation}
where $\gamma$ is a geometrical factor taking into account the shape of the mechanical mode, which is of the order of 1. $B_{\parallel}$ is the in-plane magnetic field, considering a CNT oscillating out-of-plane. $l$ is the length of the suspended CNT, and $x_{ZPF}$ is the zero-point fluctuations of the CNT mechanical resonator.

Importantly, the critical current of a Josephson junction made out of a CNT is typically of 1-30 nA, which means that the current $i_{ZPF}$ in the cavity central conductor associated to a single photon in a standard geometry can switch the SQUID to its resistive state. To prevent the switching of the SQUID junctions, one can either reduce $i_{ZPF}$ by increasing the impedance of the bare cavity, or position the SQUID away from the current anti-node of the cavity to reduce the microwave current perceived by the SQUID junctions. Combining these two strategies, and using the model developed in \cite{bourassa2012josephson}, one obtains a single-photon optomechanical coupling $g_0 = 1\,$MHz, while maintaining a critical photon number $n_C \simeq 10$. For this estimation, we used the following parameter values:

\begin{align*} 
\text{Cavity length} &=  1\,\text{mm} \\ 
\text{Cavity resonance frequency} &=  4\,\text{GHz} \\
\text{Cavity total impedance} &=  960\,\Omega \\
\text{Junction critical current} &=  10\,\text{nA} \\
\text{SQUID position} &=  0.9 \times \frac{\lambda}{4}\\
B_{\parallel} &=  200\,\text{mT} \\
\text{CNT length} &=  1\,\mu\text{m} \\
x_{ZPF} &=  1\,\text{pm} \\
\phi_S &=  0.4 \\
\end{align*}

\bibliographystyle{unsrt}
\bibliography{Supplementary_material}